\documentclass[aps,prl,floatfix,twocolumn,showpacs]{revtex4}
\usepackage{graphicx} \usepackage{amsmath} \usepackage{amssymb}
\newcommand{\BEQ}{\begin{equation}}
\newcommand{\EEQ}{\end{equation}}
\newcommand{\BEA}{\begin{eqnarray}}
\newcommand{\EEA}{\end{eqnarray}}

\renewcommand{\d}{{\rm d}}

\newcommand{\eps}{\varepsilon}

\newcommand{\A}{{\rm A}}
\newcommand{\B}{{\rm B}}
\newcommand{\C}{{\rm C}}
\renewcommand{\c}{C}

\renewcommand{\H}{{\cal H}}

\newcommand{\lb}{\langle\,}
\newcommand{\rb}{\,\rangle}
\newcommand{\tila}{\widetilde{\lambda}}
\newcommand{\tir}{\widetilde{r}}

\newcommand{\la}{\lambda}
\newcommand{\om}{\omega}

\newcommand{\lapr}{\lambda_{pr}}

\newcommand{\vb}{\vert\,}

\begin{document}
\draft
\title
{Quantum restrictions on transfer of matrix elements}
%%\date{\today}
\author{Armen E. Allahverdyan$^1$ and Karen Hovhannisyan$^2$}
\affiliation{$^1$Yerevan Physics Institute,
Alikhanian Brothers Street 2, Yerevan 375036, Armenia,\\
$^2$Yerevan State University, A. Manoogian Street 1, Yerevan, Armenia}

\begin{abstract} 

We discuss restrictions imposed by quantum mechanics on the process of
matrix elements transfer from the one system to another.  This is
relevant for various processes of partial state transfer 
(quantum communication, indirect measurement, polarization
transfer, {\it etc}).  Given two systems $\A$ and $\B$ with initial
density operators $\lambda$ and $r$, respectively, we consider most
general interactions, which lead to transferring
certain matrix elements of unknown $\lambda$ into those of the final
state ${\widetilde r}$ of $\B$.  We find that this process leads to
eliminating the memory on the transferred (or certain other) matrix
elements from the final state of $\A$.  If one diagonal matrix element
is transferred: ${\widetilde r}_{aa}=\lambda_{aa}$, the memory on each
non-diagonal element $\lambda_{a\not=b}$ is completely eliminated from
the final density operator of $\A$.  The transfer of a non-diagonal
element: ${\widetilde r}_{ab}=\lambda_{ab}$ eliminates the memory on the
diagonal elements $\lambda_{aa}$ and $\lambda_{bb}$, while the memory
about their sum $\lambda_{aa}+\lambda_{bb}$ is kept.  Moreover, the
memory about $\lambda_{ab}$ itself is completely eliminated from the
final state of $\A$.  Generalization of these set-ups to non-ideal
transfer brings in a trade-off between the quality of the transfer and
the amount of preserved memory. This trade-off is expressed via system-independent
uncertainty relations.

\end{abstract}

\pacs{03.67.-a, 03.65.-w}

%%03.65.-w Quantum mechanics 
%% 03.67.-a Quantum information; 

\maketitle

Partial or complete transfer of a quantum state from one system to
another is an essential part of many processes of energy and/or
information/entropy transport: {\it i)} Quantum communication via
(partial) state transfer plays an important role both for practical
implementation of scalable quantum processors and for understanding the
efficiency of quantum computation; see
\cite{state_transfer_burg,state_transfer_chin} for reviews.  {\it ii)}
Quantum measurements, where the initial probabilities of an observable
$\hat{A}$ of the tested system are mapped to the final probabilities of
an apparatus observable $\hat{B}$ \cite{meas}. For instance, the quantum
state of the readout object (e.g., qubit) is partially transferred to a
macroscopic system, ensuring its reliable registration \cite{kay}.  {\it
iii)} Polarization transfer from one system to another is well known in
NMR/ESR, quantum/atomic optics, semiconductor physics {\it etc}
\cite{soren, exp_2,cooling}. This is the main method of polarization
increasing or cooling \cite{cooling}.  {\it iv)} Related processes of energy (excitation) transfer
are important in biological systems (e.g., photosynthesis)
\cite{scholes}.

Here we study fundamental restrictions imposed by quantum mechanics on
information transmission via partial state transfer. To formulate this
problem, we assume that the information is encoded into matrix elements
of the system density operator (state); this situation is realized in
many of the above examples. To be a carrier of information this state
has be unknown for the transmitter, and (as the first step) we assume
that the state is {\it completely} unknown.  Now as far as the full
state transfer is concerned, one limitation comes from the no-cloning
theorem, which states that once the full (unknown) state is transferred
from system A to system B, the final state of A must differ from its
initial state \cite{wz}. Together with its various generalizations
\cite{yu,brd} the no-cloning theorem is one of the most known
constraints on the quantum information processing.  However, it cannot
be applied directly to the present problem, since here only certain (in
general, not all) matrix elements are transferred.

Consider a quantum system
$\A$ in an unknown state described by a density operator $\la$ and a
composite system $\B+\C$ in some known state with density operator
$\om$. The Hilbert spaces of $\A$ and $\B$ have the same dimension: ${\rm
dim}\H_\A={\rm dim}\H_\B=N$. The initial state of the overall system 
$\A+\B+\C$ is $\la\otimes\om$. Let $p,r=1,\ldots,N$ and
\BEA 
\label{1}
\{|p\rb\}_{p=1}^n,\,\, \lb p\,|\,r\rb=\delta_{pr},\,\,\,\,\,\,
\{|\bar{p}\rb\}_{p=1}^n,\,\, \lb
\bar{p}\,|\,\bar{r}\rb=\delta_{pr}, 
\EEA
be two orthonormal bases in $\H_\A$ and $\H_\B$, respectively. 
The interaction between $\A$ and $\B+\C$ is described by unitary
operator $U$. It will be chosen such that 
for {\it any} initial density operator $\la$ of $\A$, certain
initial matrix elements $\la_{ab}=\langle a|\la|b\rangle$ of $\rho$ are
equal to the corresponding matrix elements of the final state $\tir$ of
$\B$: 
\BEA
\la_{ab}=
\tir_{ab}=\langle \bar{a}|\tir|\bar{b}\rangle, \qquad \tir={\rm
tr}_{\A+\C} (U\,\la\otimes\om\, U^\dagger).\nonumber
\EEA 
Here $\C$ is an auxiliary system (ancilla), which ensures most
general operations.  We aim to understand implications of the
matrix elements transfer from $\A$ to $\B$ on the memory of the
transferred elements $\la_{ab}$ (or some other elements of $\la$) in the
final state $\tila={\rm tr}_{\B+\C} (U\,\la\otimes\om\, U^\dagger)$ of
$\A$. Take as an example two spin-$\frac{1}{2}$ density matrices for A and B, respectively:
$\la=\frac{1}{2}[1+\vec{l}\,\vec{\sigma}]$,
$\rho=\frac{1}{2}[1+\vec{r}\,\vec{\sigma}]$, where $\vec{\sigma}$ are
Pauli matrices, and $\vec{l},\, \vec{r}$ are Bloch vectors.  
Transferring diagonal (non-diagonal) elements $\la_{11}=\tir_{11}$
($\la_{12}=\tir_{12}$) amounts to transferring the $z$ ($x$ and $y$)
component(s) of the Bloch vectors. Both these processes are well-studied
experimentally \cite{exp_1,soren,exp_2,cooling}. 

The initial state of $\B+\C$ is chosen as 
\BEA
\om=|\,\bar{1}\rb\otimes|\,\c \rb\,\lb \bar{1}\,|\otimes\lb \c \,|,
\label{ini} 
\EEA 
where $|\c \rangle$ lives in the Hilbert space $\H_\C$ of $\C$. This
choice does not restrict generality provided that we are free to choose
the system $\C$ and design unitary evolutions for $\B+\C$. Indeed, an
initial mixed state of $\B+\C$ can be purified by extending $\C$ to a
larger Hilbert space, while the resulting pure state can be rotated to
$|\,\bar{1}\rb\otimes|\,\c\rb$ by a suitable unitary operator. 

We represent the unitary operator $U$ as ($p=1,\ldots,N$)
\BEA
\label{riogo}
\label{unitar} U\,|p\rb\otimes |\,\bar{1}\rb\otimes|\,\c \rb ={\sum}_{k,l}
|k\rb\otimes|\,\bar{l}\rb\otimes|\,\c^{p}_{kl}\rb\equiv |\psi_p\rangle, 
\EEA 
where all summation indices run from $1$ to $N$, and where the vectors
$|\,\c^{p}_{kl}\rb$ with $p,k,l=1,\ldots,N$ live in $\H_\C$.

The unitarity of $U$ amounts to ($p,r=1,\ldots,N$)
\BEA
\label{bao}
\langle\psi_p|\psi_r\rangle=\delta_{rp}\qquad {\rm or}\qquad
{\sum}_{kl}\lb \c^p_{kl}\,|\,\c^{r}_{kl}\rb=\delta_{rp}.
\EEA
Clearly, (\ref{riogo}, \ref{bao}) define for our purposes the most general
unitary operation. 
The final states $\tila$ and $\tir=\sum_{a,b} r_{ab}|\bar{a}\rb\lb\bar{b}|$
of $\A$ and $\B$, respectively, read from (\ref{unitar})
\BEA
\label{kh1} \tila={\sum}_{pr} \lapr \Theta_{rp},~
\Theta_{rp}\equiv {\sum}_{kn} |k\rb\,\lb n|{\sum}_{l}\lb
\c^r_{nl}|\,\c^p_{kl}\rb, \\
\tir_{ab} =
{\sum}_{pr}\lapr{\sum}_{k} \lb \c^r_{kb}|\,\c^p_{ka}\rb.
\label{kh2} 
\EEA
The process of matrix elements transfer depends crucially on which
(diagonal or non-diagonal) elements are transferred. In a sense,
diagonal (non-diagonal) elements represent classical (quantum) aspects
of the information contained in the unknown state $\la$. In particular,
the transfer of non-diagonal elements relates to transferring
entanglement. We therefore study these cases separately. 

{\it Diagonal to diagonal transfer.} 
Assume that for every initial state $\la$ of $\A$ a diagonal element $\lambda_{aa}$ of $\A$ is transferred
into the diagonal element $\tir_{aa}$ of B: $\lambda_{aa}= \tir_{aa}$. 
For this it is necessary to have [see (\ref{kh2})]
\BEA 
\label{ba}
{\sum}_k\lb \c^r_{ka}|\,\c^p_{ka}\rb=\delta_{pr}\delta_{pa}
~~{\rm for ~ all~pairs}~(r,p). 
\EEA
Eq.~(\ref{ba}) for $r=p=a$ implies 
${\sum}_k\lb \c^a_{ka}|\,\c^a_{ka}\rb=1$. Combining this with
(\ref{bao}) under the same condition $p=r=a$ gives
$|\c^a_{kl}\rb=0$ for $l\not=a$. Eq.~(\ref{ba}) for $r=p=c\not=a$ gives
$|\c^c_{ka}\rb=0$ for every $c\not=a$. Altogether, we get
\BEA
{\sum}_l\lb \c^a_{nl}\,\vb
\c^c_{kl}\rb=0~~{\rm for~every}~c\not=a~~{\rm or}~~\Theta_{ac}=0,\nonumber
\label{tu3} 
\EEA 
implying from (\ref{kh1}) that {\it due to the transfer $\lambda_{aa}=
\tir_{aa}$ the memory on each initial non-diagonal element
$\lambda_{a\not =c}$ in the final density operator $\tila$ of $\A$ is
lost}. This generalizes the no-cloning principle, since once the memory
of some elements is eliminated from the final state of $\A$, this state
cannot be kept intact.  Note that {\it i)} to be able to speak on the
memory and its loss, we have to have initially some freedom in choosing
$\lambda_{a\not =c}$, i.e., the latter should carry some information.
{\it ii)} $\tila$ need not be diagonal. {\it iii)} The memory on $\la_{aa}$
itself is conserved in $\tila$. 

While the above results refer to the ideal transfer, it is important to
see how much memory can be preserved under a non-ideal transfer. The
simplest definition of the non-ideal transfer for one matrix element
$\lambda_{aa}$ amounts to requiring
$\tir_{aa}=\varepsilon_a\,\lambda_{aa}$, where $0<\varepsilon_a<1$ does
not depend on the initial $\la$ and quantifies the non-ideality. Thus,
if $\lambda_{aa}$ is considered as signal, $\varepsilon<1$ corresponds
to reducing (by a fixed amount) the signal magnitude. If some noise is
present during the transfer, this reduction will correspond to
decreasing the signal-to-noise ratio. 

We now study the {\it maximal possible memory} on the initial
non-diagonal elements $\la_{a\not=c}$ under such transfer. It proves
more convenient to assume $N\geq 3$ and to start immediately with the
simultaneous non-ideal transfer of two diagonal elements:
\BEA\label{dd} 
\tir_{aa}=\varepsilon_a\,\lambda_{aa}, ~~\tir_{bb}=\varepsilon_b\,\lambda_{bb}, ~~0<\varepsilon_a<1,
~~0<\varepsilon_b<1,
\EEA
where $\varepsilon_a$ and $\varepsilon_b$ do not depend on the initial
state $\la$ and quantify the non-ideality.  This case is generic, since
the non-ideal transfer of one (or several) elements can be recovered
from it; see below. (The non-ideal transfer (\ref{dd}) does not exist for
$N=2$, since the trace should be conserved; here we can transfer only one element.)
Instead of (\ref{ba}) we get from (\ref{dd})
\BEA
\label{aba}
{\sum}_k\lb \c^r_{ku}|\,\c^p_{ku}\rb=\varepsilon_u\delta_{pr}\delta_{pu}
~{\rm for~ all}~ (r,p)~{\rm and}~u=a,b.
\EEA
Eq.~(\ref{aba}) for $r=p\not=a$ and for $r=p\not=b$ gives for any $k$
\BEA
\label{babo}
|\c^p_{ka}\rb=0~~ {\rm for}~ p\not=a~~ {\rm and}~~
|\c^p_{kb}\rb=0~~ {\rm for}~ p\not=b.
\EEA
The memory of the final state (\ref{kh1}) on the non-diagonal element
$\la_{a\not= c}$ should be quantified via derivative of $\tila$ over
$\la_{a\not= c}$. Though due to the presence of $\la_{a\not= c}^*$,
$\tila$ is not an analytic function of $\la_{a\not= c}$, we can employ
the generalized complex-variable derivative \cite{haykin}:
\BEA
\frac{\partial\tila}{\partial \la_{a c}}
\equiv \frac{1}{2}\left(\left.\frac{\partial\tila}{\partial \,\Re\la_{a c}}\right|_{\Im\la_{ac}}
-i\left.\frac{\partial\tila}{\partial \,\Im\la_{ac}}\right|_{\Re\la_{ac}}
\right)=\Theta_{ac},
\label{loran}
\EEA
where $\Theta_{ac}$ is defined in (\ref{kh1}).  The definition
(\ref{loran}) has all features expected from a derivative \cite{haykin}.
In particular, for an analytic (over $\la_{a\not=c}$) function it
coincides with the ordinary complex derivative \cite{haykin}. 

The magnitude of the
matrix $\Theta_{a\not=c}$, or the strength of the dependence of $\tila$
on $\la_{a\not=c}$, can be characterized by some norm. Since all norms
are equivalent in a finite-dimensional Hilbert space \cite{haykin}|i.e.,
given two norms $||.||_1$ and $||.||_2$, there exist positive constants
$a$ and $b$ such that $a||A||_2 \leq ||A||_1 \leq b ||A||_2$ for any
matrix $A$|we work with the Euclidean norm $||\Theta_{ac}||\equiv
\sqrt{{\rm tr}(\Theta_{ac}\Theta^\dagger_{a c})}$, where
$\Theta^\dagger$ is the hermitean conjugate of $\Theta$. Recall that
for any norm $||A||=0$ implies $A=0$. Due to (\ref{loran}) we get
\BEA
||\Theta_{a\not=c}||=\frac{1}{2}
\sqrt{||{\partial\tila}/{\partial \,\Re\la_{a c}}||^2 +
||{\partial\tila}/{\partial \,\Im\la_{a c}}||^2}, 
\EEA
showing that $||\Theta_{a\not=c}||$ includes the memory on the real and imaginal part of $\tila$.
The same value
$||\Theta_{a\not=c}||$ is obtained under norming the complex conjugate
derivative ${\partial\tila}/{\partial \la^*_{a c}}= (\,
{\partial\tila^\dagger}/{\partial \la_{ac}}\,)^\dagger=
\Theta^\dagger_{ac}$. 

That the memory of $\tila$ on $\la_{ac}$ can be characterized by
$||\Theta_{ac}||$ is verified also by studying the matrix gradient of
$\tila$, whose modulus is limited by $||\Theta_{a\not=c}||$ and
$\frac{1}{\sqrt{2}}||\Theta_{a\not=c}||$ from above and below,
respectively \cite{gradient}. Note that in the initial state
$||\frac{\partial\la}{\partial \la_{ac}}||=1$, while in general
$||\Theta_{ac}||\leq 1$; see (\ref{grom1}--\ref{grom3}) below. Thus,
expectedly, the memory on matrix element can only decrease after a
unitary transformation. 

Given (\ref{dd}, \ref{aba}, \ref{babo}) we now establish an upper bound on $||\Theta_{a\not=c}||$.  Let
$z^{r~p}_{nl\,kl}\equiv \lb \c^r_{nl}|\,\c^p_{kl}\rb$ and let ${\sum}'_{l}$
be the summation over $l=1,\ldots,N$ excluding $l=a$ and $l=b$. We get
from (\ref{kh1})
\BEA
\label{grom1}
||\Theta_{a\not=c}||^2\equiv {\sum}_{k,n}\left|{\sum}_l z^{c~a}_{nl\,kl}\right|^2
\leq {\sum}_{k,n}\left[{\sum}_l |z^{c~\,a}_{nl\,kl}|\right]^2\\
\label{grom2}
\leq {\sum}_{k,n}\left[{\sum}'_l \sqrt{z^{c~\,c}_{nl\,nl}}\sqrt{z^{a~a}_{kl\,kl}}  \right]^2~~~\\
\leq {\sum}_n{\sum}'_l z^{c~\,c}_{nl\,nl}\,\,\, {\sum}_k{\sum}'_l z^{a~a}_{kl\,kl},~~~
\label{grom3}
\EEA
where the inequalities in (\ref{grom2}) and (\ref{grom3}) are due to the Cauchy-Schwartz
inequality, while in (\ref{grom2}) we additionally used (\ref{babo}). We 
now get from (\ref{grom3}) and (\ref{bao}, \ref{aba}, \ref{babo})
\BEA
\label{odzin1}
&&||\Theta_{a\not=b}||\leq \sqrt{(1-\varepsilon_a)(1-\varepsilon_b)},~~~~\\
&&||\Theta_{a\not=c}||\leq \sqrt{(1-\varepsilon_a)}~~{\rm for~every}~~ c\not =a,\, c\not =b.~~~~
\label{odzin2}
\EEA
These inequalities|which are akin to the uncertainty relations|relate
non-ideality of the transfer to the maximal possible amount of the
conserved memory. The extension of
(\ref{odzin1}, \ref{odzin2}) to transferring non-ideally several matrix
elements should be obvious, since the non-diagonal elements under such a
transfer divide naturally into two classes, which correspond to
(\ref{odzin1}) and (\ref{odzin2}), respectively. 

Let us show that the bounds (\ref{odzin1}, \ref{odzin2}) are
saturated by the proper choice of $|\c^p_{kn}\rb$. To this end assume
that ${\rm dim}\H_\C=1$: $|\,\c^p_{kb}\rb=\c^p_{kb}|\,\c\rb$, where $\c^p_{kb}$
are c-numbers satisfying (\ref{bao}). Choosing for $N=3$
\BEA
\c^1_{11} = \sqrt{\varepsilon_1},~
\c^1_{13} = \sqrt{1-\varepsilon_1},~
\c^2_{22} = \sqrt{\varepsilon_2},~
\c^2_{23} = \sqrt{1-\varepsilon_2},\nonumber
\EEA
and $\c^3_{33} = 1$ (while all other $\c^p_{kb}$ with $p,k,b=1,2,3$ are zero) we satisfy the
unitarity conditions (\ref{bao}) and realize the optimal
memory-conserving non-ideal transfer (\ref{dd}) with $a=1$ and $b=2$.
Now (\ref{odzin1}, \ref{odzin2}) become equalities. 

The memory on the transferred diagonal elements $\la_{uu}$ ($u=a,b$) in
the final state $\tila$ of $\A$ is quantified by the norm
$||\Theta_{uu}||\leq 1$. The above example is optimal
with respect to the memory-conservation of the non-diagonal elements,
and it also provides the maximal memory of the transferred
elements: $||\Theta_{11}||=||\Theta_{22}||=1$. 

{\it Nondiagonal to nondiagonal transfer.} Demanding 
\BEA
\label{karait1}
{\sum}_k \lb \c^r_{ka}|\,\c^p_{kb}\rb=\delta_{ra}\delta_{pb}~~{\rm for ~ all}~(r,p)
~{\rm and}~a\not=b,
\EEA
amounts to transferring ideally the corresponding non-diagonal element:
$\tir_{ab}=\lambda_{ab}$ for arbitrary initial state $\la$ of $\A$; see (\ref{kh1}). 
The non-negativity of 
${\sum}_k[\alpha^*\lb \c^a_{ka}|+\beta^*\lb \c^b_{kb}| \,]\,[\,\alpha|\,\c^a_{ka}\rb+\beta|\,\c^b_{kb}\rb]$
as a function of two complex numbers $\alpha$ and $\beta$ (Cauchy-Schwartz inequality) leads to
\begin{gather}
1={\sum}_k\lb \c^a_{ka}|\,\c^b_{kb}\rb\leq \sqrt{{\sum}_k \lb \c^a_{ka}|\,\c^a_{ka}\rb
{\sum}_k \lb \c^b_{kb}|\,\c^b_{kb}\rb },\nonumber\\
\label{jan}
\end{gather}
where the equality in (\ref{jan}) is due to (\ref{karait1}) under $r=a$ and $k=b$. The
inequality in (\ref{jan}) has to be saturated, since (\ref{bao}) implies
${\sum}_k\lb \c^a_{ka}|\,\c^a_{ka}\rb\leq 1$, ${\sum}_k\lb \c^b_{kb}|\,\c^b_{kb}\rb\leq 1$.
Thus we have ${\sum}_k\lb \c^a_{ka}|\,\c^a_{ka}\rb={\sum}_k\lb \c^b_{kb}|\,\c^b_{kb}\rb= 1$,
which together with (\ref{bao}) gives for any $k$
\BEA
\label{dedo}
|\c^a_{kl}\rb=0~~ {\rm for}~ l\not=a~~ {\rm and}~~
|\c^b_{kl}\rb=0~~ {\rm for}~ l\not=b.
\EEA
Eqs.~(\ref{kh1}, \ref{dedo}) lead to $\Theta_{a\not=b}=\Theta_{b\not=a}=0$, i.e.,
{\it the memory on the transferred non-diagonal element
$\lambda_{ab}$ in the final density operator $\tila$ is lost}.

Another consequence of saturating the inequality in (\ref{jan}) is that
$|\c^b_{kb}\rb=|\c^a_{ka}\rb$ for any $k$, which leads to
\BEA
{\sum}_l\langle \c^a_{nl}|\c^a_{kl}\rangle
=\langle \c^a_{na}|\c^a_{ka}\rangle
= {\sum}_l\langle \c^b_{nl}|\c^b_{kl}\rangle
=\langle \c^b_{nb}|\c^b_{kb}\rangle,\nonumber
\EEA
i.e., $\Theta_{aa}=\Theta_{bb}$, meaning that {\it the memory about the
diagonal elements $\lambda_{aa}$ and $\lambda_{bb}$ in the final density
operator $\tila$ is lost.  Only the memory about
$\lambda_{aa}+\lambda_{bb}$ is kept}. Thus one ideal
nondiagonal-to-nondiagonal transfer eliminates the memory on three real
quantities, while one diagonal-to-diagonal ideal transfer eliminates
memory on $2(N-1)$ real quantities. The difference between these two
cases is that for the ideal nondiagonal-to-nondiagonal transfer the
memory on the transferred element itself is eliminated from the final state
of $\A$. Let us announce that when only the real part of the non-diagonal
element is transferred, $\Re\,\tir_{ab}=\Re\,\lambda_{ab}$ for any $\la$, then the above
result on eliminating the memory on $\lambda_{aa}$ and $\lambda_{bb}$ still
holds, while only the memory on the imaginary part $\Im\lambda_{ab}$ is
eliminated from the final density operator $\tila$ of $\A$ \cite{k}. 

Turning to the non-ideal transfer $\tir_{ab}=\eps\lambda_{ab}$ (where
$a\not =b$ are two indices and $0<\eps<1$), we restrict ourselves to
finding the maximal possible value of $||\Theta_{12}||$ for the c-number
case $|\c^p_{kb}\rb=\c^p_{kb}|\,\c\rb$, since so far we were not able to
get more general results. For $\tir_{a\not=b}=\eps\lambda_{a\not=b}$ to
hold for arbitrary initial state $\la$ of $\A$ we need
\BEA
\label{varaz}
{\sum}_k \c^r_{kb}\c^{p\,\,*}_{ka}=\eps\delta_{rb}\delta_{pa}~~{\rm for ~ all}~(r,p)
~{\rm and}~a\not=b.
\EEA
This implies ${\sum}_k \c^a_{kb}\c^{a\,\,*}_{ka}={\sum}_k \c^b_{kb}\c^{b\,\,*}_{ka}=0$,
and then
\begin{gather}
\label{kom1}
||\Theta_{a\not=b}||^2
=\phi^{a}_{a}\phi^{b}_{a}+\phi^{a}_{b}\phi^{b}_{b}+\Lambda_{12},~
\phi^{u}_{v}\equiv {\sum}_k |\c^u_{kv}|^2,  \\
\Lambda_{a\not=b}\equiv {\sum}'_{[sl]}\left[{\sum}_k \c^a_{kl}\c^{a\,\,*}_{ks}\right]
\left[{\sum}_n \c^b_{ns}\c^{b\,\,*}_{nl}\right],
\label{kom2}
\end{gather}
where ${\sum}'_{[sl]}$ means that the values $(s,l)=(a,a), (a,b), (b,a), (b,b)$
are excluded from the summation over $s=1,\ldots,N$ and $l=1,\ldots,N$.
In estimating $|\Lambda_{a\not=b}|$ from above we proceed by applying the Cauchy-Schwartz inequality
and using (\ref{kom1}):
\BEA
|\Lambda_{a\not=b}|\leq 
{\sum}'_{[sl]}\left[{\sum}_k |\c^a_{kl}||\c^{a\,\,*}_{ks}|\right]
\left[{\sum}_n |\c^b_{ns}||\c^{b\,\,*}_{nl}|\right]\\
\leq {\sum}'_{[sl]} \sqrt{ \phi^{a}_{l} \phi^{a}_{s} \phi^{b}_{l} \phi^{b}_{s}   }
\leq \sqrt{  
{\sum}'_{[sl]}\phi^{a}_{l} \phi^{a}_{s} {\sum}'_{[sl]}\phi^{b}_{l} \phi^{b}_{s}
}.
\label{dodosh3}
\EEA
Working out (\ref{dodosh3}) and combining it with (\ref{kom1}) we obtain
\BEA
\label{bern1}
||\Theta_{a\not=b}||^2
&\leq& \phi^{a}_{a}\phi^{b}_{a}+\phi^{a}_{b}\phi^{b}_{b}~~~~~~
\\
&+&
\sqrt{[1-(\phi^a_a+\phi^a_b)^2]  [1-(\phi^b_a+\phi^b_b)^2]  }\equiv F.~~~~~
\label{bern2}
\EEA
We now maximize $F$ in the RHS of (\ref{bern2}) so as to obtain a bound
on $||\Theta_{a\not=b}||^2$ that holds for any $\{\c^b_{kl}\}$. The
maximization is carried out under two constraints: {\it i)}
$\phi_a^a\phi_b^b\geq \eps^2$, which follows from applying the
Cauchy-Schwartz inequality to (\ref{varaz}) with $r=b$ and $p=a$; {\it
ii)} $\phi_a^a+\phi_b^a\leq 1$ and $\phi_b^b+\phi_a^b\leq 1$, which
follow from the unitarity condition (\ref{bao}). Note from (\ref{bern1}, \ref{bern2})
that the maximum of $F$ over $\phi^b_a$ can be reached only at the
boundaries of its range, i.e., at $\phi^b_a=0$ or at
$\phi^b_a=1-\phi_b^b$.  The same holds for $\phi^a_b$. Direct
inspection shows that the maximum of $F$ is reached for
$\phi^b_a=\phi^a_b=0$ and $\phi_a^a=\phi_b^b=\eps$:
\BEA
\label{opus}
||\Theta_{a\not=b}||\leq\sqrt{1-\eps^2}.
\EEA
Comparing (\ref{opus}) with (\ref{odzin2}) we see that the
maximal amount of the preserved memory for the non-ideal
nondiagonal-to-nondiagonal transfer is larger than for the non-ideal
diagonal-to-diagonal transfer. 

For the transfer $\tir_{12}=\eps\lambda_{12}$ and for $N=2$
the bound (\ref{opus}) is saturated by the following choice of 
$\{\c^b_{kl}\}$
\BEA
&&\c^1_{11} = 1,~\c^1_{21}=\c^1_{12}=\c^1_{22} = 0,\\
&&\c^2_{21} = \sqrt{1-\eps^2},~ \c^2_{12} = \eps, ~\c^2_{22}=\c^2_{11}=0.
\EEA
Generalizing this example to $N\geq 3$ is straightforward.

{\it Two non-commuting initial states.} Above we assumed that the
initial state $\la$ of the (source) system $\A$ is completely unknown.
Let us now assume that due to some {\it a priori} information $\la$ can
be one of two non-commuting density operators $\rho$ and $\chi$. This
is the minimal setup, which still contains quantum information
\cite{brd}.  The complete understanding of the matrix elements transfer
for this minimal setup is yet to be developed \cite{k}. Here we present
one example showing that the above constraints on the transfer may or
may not carry over literally.  Assume that $\rho$ and $\chi$ are two
$2\times 2$ matrices with the following relation between their matrix
elements:
\BEA
\rho_{11}\,\chi_{12}=\chi_{11}\,\rho_{12},~~
\rho_{11}\not =\chi_{11},~~
\rho_{12}\not=\chi_{12},
\label{castro}
\EEA
where $\rho$ and $\chi$ do not commute due to the latter two conditions.
The relations for the exact transfer of the 
diagonal elements $\rho_{11}$ and $\chi_{11}$ [see (\ref{kh2}) with $N=2$] read:
$\rho_{11}={\sum}_{p,r=1}^2\,\rho_{pr}{\sum}_{k=1}^2 \lb \c^r_{k1}|\,\c^p_{k1}\rb$,
$\chi_{11}={\sum}_{p,r=1}^2\,\chi_{pr}{\sum}_{k=1}^2 \lb \c^r_{k1}|\,\c^p_{k1}\rb$.
(Then $\rho_{22}$ and $\chi_{22}$ are
transferred automatically.) Multiplying the first and second equation by
$\chi_{12}$ and $\rho_{12}$, respectively,
subtracting the resulting equations from each other and employing
(\ref{castro}) and (\ref{bao}) we conclude that for $k=1,2$:
$|\c^2_{k1}\rangle=|\c^1_{k2}\rangle=0$. Together with (\ref{kh1}) this
leads to eliminating the memory on the non-diagonal elements:
$\Theta_{12}=0$. However, under conditions (\ref{castro}) the exact
transfer of the non-diagonal element does not lead to any complete elimination of
memory. 

{\it In conclusion}, we studied how quantum mechanics constrains the
process of matrix elements transfer from one system $\A$ to another
system $\B$. Assuming that the initial state of $\A$ is completely
unknown, we show that transferring certain matrix elements leads to
eliminating the memory on the transferred (or certain other) matrix
elements from the final state of $\A$. We also studied the maximal
memory that can be preserved under non-ideal transfer. For each type of
transfer this maximal memory relates to the amount of non-ideality by
universal relations akin to the uncertainty relations. 

It is pleasure to thank R. Balian for discussions.
The work was supported by Volkswagenstiftung.
%grant ``Quantum Thermodynamics: Energy and information flow at nanoscale''.


\begin{thebibliography}{99}
\bibitem{gradient}Consider the differential $\d\tila$ of the matrix $\tila$ over an infinitesemal
increments of two independent real variables $x\equiv\Re\la_{a\not=c}$ and
$y\equiv\Im\la_{a\not=c}$: $\d\tila=\Theta_{ac}(\d x+i\d y)+\Theta^\dagger_{ac}(\d x-i\d y)$.
To develop the notion of the matrix gradient (similar to the usual gradient)
we first calculate the Euclidean norm $||\d\tila||$ of $\d\tila$:
$||\d\tila||^2=(\d x^2+\d y^2)[{\rm tr}(A^2) n_x^2+{\rm tr}(B^2) n_y^2+2{\rm tr}(A B) n_x n_y]$,
where $A=\Theta_{ac}+\Theta_{ac}^\dagger$, $B=i(\Theta_{ac}-\Theta_{ac}^\dagger)$, and where
$(n_x,n_y)=( \d x,\d y)/\sqrt{\d x^2+\d y^2}$ is the normal
vector over the direction of the small increment. The (modulus of) matrix gradient 
is defined by maximizing over $\vec{n}\equiv (n_x,n_y)$: $\frac{1}{4}{\rm max}_{\vec{n}}
||\d\tila||^2=\frac{1}{2}(\d x^2+\d y^2)[\,{\rm tr}( \Theta_{ac}\Theta_{ac}^\dagger )
+|{\rm tr}( \Theta_{ac}^2)|\,]$. 
As follows from ${\rm tr}(A B)\leq\sqrt{{\rm tr}(A^2){\rm tr}(B^2)}$, 
$\frac{{\rm max}_{\vec{n}}
||\d\tila||^2}{4(\d x^2+\d y^2)}$ is limited from above by 
${\rm tr}( \Theta_{ac}\Theta_{ac}^\dagger )=||\Theta_{ac}||^2$, and from below
by $\frac{1}{2}{\rm tr}( \Theta_{ac}\Theta_{ac}^\dagger )$.
\end{thebibliography}
\end{document}